\documentclass[doublecol]{epl2}

\title{Stable half-metallic ferromagnetism in nonstoichiometric cubic binary chromium chalcogenides}
 \shorttitle{Stable half-metallic ferromagnetism in chromium chalcogenides}

\author{San-Dong Guo \and Bang-Gui Liu\thanks{E-mail: \email{bgliu@mail.iphy.ac.cn}}}
\shortauthor{San-Dong Guo and Bang-Gui Liu}

\institute{ Institute of Physics, Chinese Academy of
Sciences, Beijing 100190, China\\
Beijing National Laboratory for Condensed Matter Physics, Beijing
100190, China}


\abstract{We find that three nonstoichiometric cubic binary
chromium chalcogenides, namely Cr$_3$S$_4$, Cr$_3$Se$_4$, and
Cr$_3$Te$_4$, are stable half-metallic ferromagnets with wide
half-metallic gaps on the basis of systematic state-of-the-arts
first-principles calculations. We optimize their structures, and
then calculate their magnetic moments, electronic structures,
formation heats, and elastic moduli and investigate their
structural stability and robustness of ferromagnetism against
antiferromagnetic fluctuations. Our calculated results show that
the three sulvanite phases are structurally stable and
ferromagnetically robust, and hence could be realized as epitaxial
thin films. We attribute the structural and ferromagnetic
stability and the better half-metallicity to their special
effective Cr valence $2.667+$. These findings will open doors for
much more high-performance spintronic materials compatible with
current semiconductor technology. }

\begin{document}

\maketitle

It is believed that next-generation high-performance computers can
be achieved through using the spin freedom of electron in key
materials and devices of current semiconductor
technology\cite{Wolf,Pickett}. The half-metallic ferromagnet, first
discovered in NiMnSb by de Groot {\it et al} in 1983, has almost
100\% spin polarization near the Fermi level\cite{hm,hm1,pickett1}.
This feature of half-metallic ferromagnetism makes carriers have
high spin-polarization near the Fermi level and avoid some
spin-related scattering processes that should exist otherwise. These
are essential to practical spintronic
applications\cite{Wolf,Pickett,pickett1}. Half-metallic
ferromagnetism has been found in many materials, such as Heusler
alloys\cite{hm,hm1,heusler}, transition-metal
oxides\cite{oxide,oxide1,oxide2}, and even graphene nanoribbons
under electric field\cite{louie}. The half-metallic ferromagnetic
materials compatible with semiconductor technology is believed to be
promising candidates for achieving more powerful computers. For this
purpose, many zincblende transition-metal pnictides and
chalcogenides\cite{nhm,cras,crsb,lbg,xie,crte,cras1,crsb1},
regularly-doped semiconductors\cite{Ohno,slj}, semiconductor
super-structures\cite{sidelta,mtj}, and more related compounds have
been proved to be half-metallic. Although great advance has been
achieved, better half-metallic ferromagnets compatible with current
semiconductor technology are highly desirable for the
next-generation computers.

Here we find three stable half-metallic ferromagnets, namely
Cr$_3$S$_4$, Cr$_3$Se$_4$, and Cr$_3$Te$_4$, among
nonstoichiometric cubic binary transition-metal chalcogenides on
the basis of state-of-the-arts first-principles calculations. We
investigate their structural stability and ferromagnetic
robustness against antiferromagnetic fluctuations. Our calculated
results show that the three nonstoichiometric sulvanite phases
have formation heats of down to -0.302 eV per formula unit with
respect to corresponding zincblende phases and their half-metallic
gaps can be up to 1.05 eV. We attribute the structural and
ferromagnetic stability and the better half-metallicity to their
special effective Cr chemical valence $2.667+$, compared with $2+$
in the stoichiometric structures. These nonstoichiometric cubic
materials have much better features for spintronic applications
based on semiconductors. More detailed results will be presented
in the following.

We use the full-potential linearized augmented-plane-waves method
within the density functional theory (DFT)\cite{dft}, as
implemented in package WIEN2k\cite{wien2k,wien2ka}. The
generalized gradient approximation (GGA)\cite{pbe96} is used for
the exchange-correlation potential. Full relativistic effects are
calculated with the Dirac equations for core states, and the
scalar relativistic approximation is used for valence states. The
spin-orbit coupling is neglected because it has little effect on
our results. We use 2000 k points in the first Brillouin zone,
make the harmonic expansion up to $l_{\rm max}$=10 in atomic
spheres, and set $R_{\rm mt}$$\times$$K_{\rm max}$ to 7.5. The
radii of the atomic spheres of Cr and others are chosen so that as
high accuracy as possible is obtained. The volumes are optimized
in terms of total energy, and the internal position parameters
with a force standard of 3 mRy/a.u. The simplest antiferromagnetic
structures are constructed by doubling the unit cells along the
[100] and [110] directions. There are six Cr atoms in each of
these doubled cells. We let three of the six Cr spins orient up
and the other three down. All the spin values of the Cr, S, Se,
and Te atoms are determined naturally by the self-consistent
calculations. The structural stability is investigated through
deforming the structures (with the volumes fixed) along the three
directions. The elastic moduli are calculated with the standard
method implemented in WIEN2k\cite{wien2k,wien2ka}. The formation
heats of the cubic sulvanite structures are calculated with
respect to the crystalline Cr phase and the corresponding
zincblende structures. The self-consistent calculations are
considered to be converged only when the absolute integrated
charge-density difference per formula unit between the two
successive loops is less than 0.0001$|e|$, where $e$ is the
electron charge.

\begin{figure}[t]
\begin{center}
\includegraphics[width=8cm]{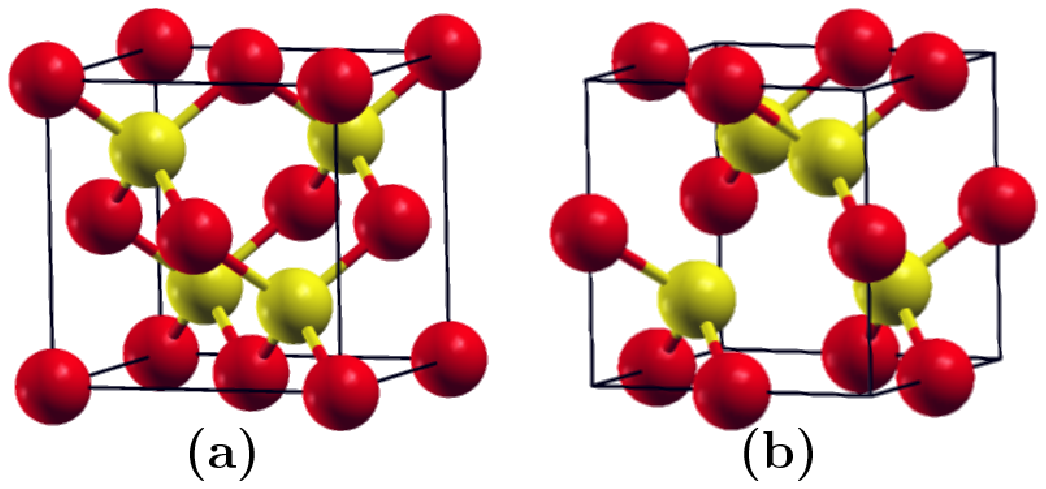}
\vspace{0mm} \caption{(color online). The crystal structures of
zincblende Cr$X$ (a) and sulvanite Cr$_3$$X_4$ (b) phases, where
$X$ can be S, Se, or Te. The red (or black) ball denotes a Cr atom
and the yellow (or gray) an $X$ atom.} \label{fig1}
\end{center}
\end{figure}

\begin{table}\begin{center}
\caption{The lattice constants $a$, the internal atomic parameters
$z_{X}$, and the Cr-$X$ bond lengths $l_v$ of the three sulvanite
Cr$_3$$X_4$ phases. Presented in parentheses are those of the
corresponding zb-Cr$X$ phases.}\label{table1}
\begin{tabular*}{0.475\textwidth}{@{\extracolsep{\fill}}cccc}
 \hline\hline
Name & Cr$_3$S$_4$ (CrS) & Cr$_3$Se$_4$ (CrSe) & Cr$_3$Te$_4$ (CrTe)\\
\hline
$a$ (\AA)   & 5.344 (5.469) & 5.679 (5.833) & 6.117 (6.292)\\
$z_{X}$     & 0.250 (0.25)  & 0.250 (0.25)  & 0.260 (0.25)\\
$l_v$ (\AA) & 2.315 (2.368) & 2.460 (2.526) & 2.686 (2.725)\\
 \hline\hline
\end{tabular*}\end{center}
\end{table}

\begin{figure}[!htbp]
\begin{center}
\includegraphics[width=8cm]{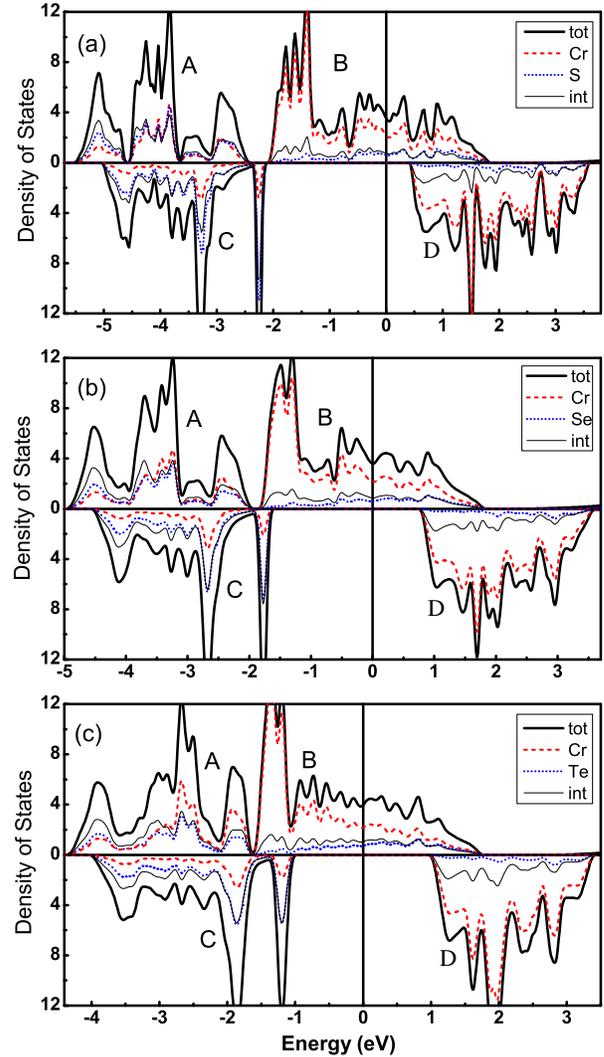}
\vspace{0mm} \caption{(color online). Spin-dependent densities of
states (DOSs) of the three sulvanite Cr$_3$$X_4$ phases for $X$ =
S (a), Se (b), and Te (c). The upper part in each of the three
panels is the majority-spin DOS and the lower part the minority
spin. The solid thick line represents the total DOS; the red (or
gray) dash, blue (or light gray) dot, and black thin lines
describe the partial DOSs projected in Cr and $X$ atom spheres and
the interstitial region, respectively. } \label{fig2}
\end{center}
\end{figure}

The ground-state phases for three stoichiometric Cr$X$ ($X$=S, Se,
and Te) compounds are of nickel-arsenide (na) structure, but the
zincblende (zb) structures of the Cr$X$ have been shown to be
higher only by 0.28-0.36 eV per formula unit than the
corresponding na ones\cite{xie}, and therefore have very good
stability. This actually stimulated experimental synthesis of
zb-CrTe epitaxial thin films with a thickness up to 100
nm\cite{crte}. The sulvanite structure of Cr$_3$$X_4$ still has
cubic symmetry with space group No. 215. Each of the cations has
four anionic neighbors, but each of the anions three cationic
neighbors. With the cations being at the ($\frac 12$,0,0) sites,
the anions occupy the ($z_X$,$z_X$,$z_X$) sites. The two cubic
structures are shown in Fig. 1. For all the sulvanite Cr$_3$$X_4$
compounds, the lattice constant $a$ and the internal structural
parameter $z_X$ are optimized fully in terms of usual total energy
and force standards, and then the Cr-$X$ bond lengths $l_v$ are
calculated with the optimized structures. We present our
calculated $a$, $z_X$, and $l_v$ results for all the three
sulvanite Cr$_3$$X_4$ in Table \ref{table1}. Those of the
corresponding zb Cr$X$ phases are shown in parentheses for
comparison. Cr$_3$S$_4$ and Cr$_3$Se$_4$ almost keep the same
$z_X$=0.25 as zb-CrS and zb-CrSe, and Cr$_3$Te$_4$ has
$z_X$=0.260, a little larger than that of zb-CrTe. The sulvanite
structures have smaller lattice constants than the zb ones by
1.6-2.6\% and smaller Cr-$X$ bond lengths by 1.4-2.6\%. In the
following, spin-dependent densities of states (DOSs), energy
bands, and charge and moment density distributions are calculated
in terms of the optimized structures.

\begin{figure}[!htbp]
\begin{center}
\includegraphics[width=8cm]{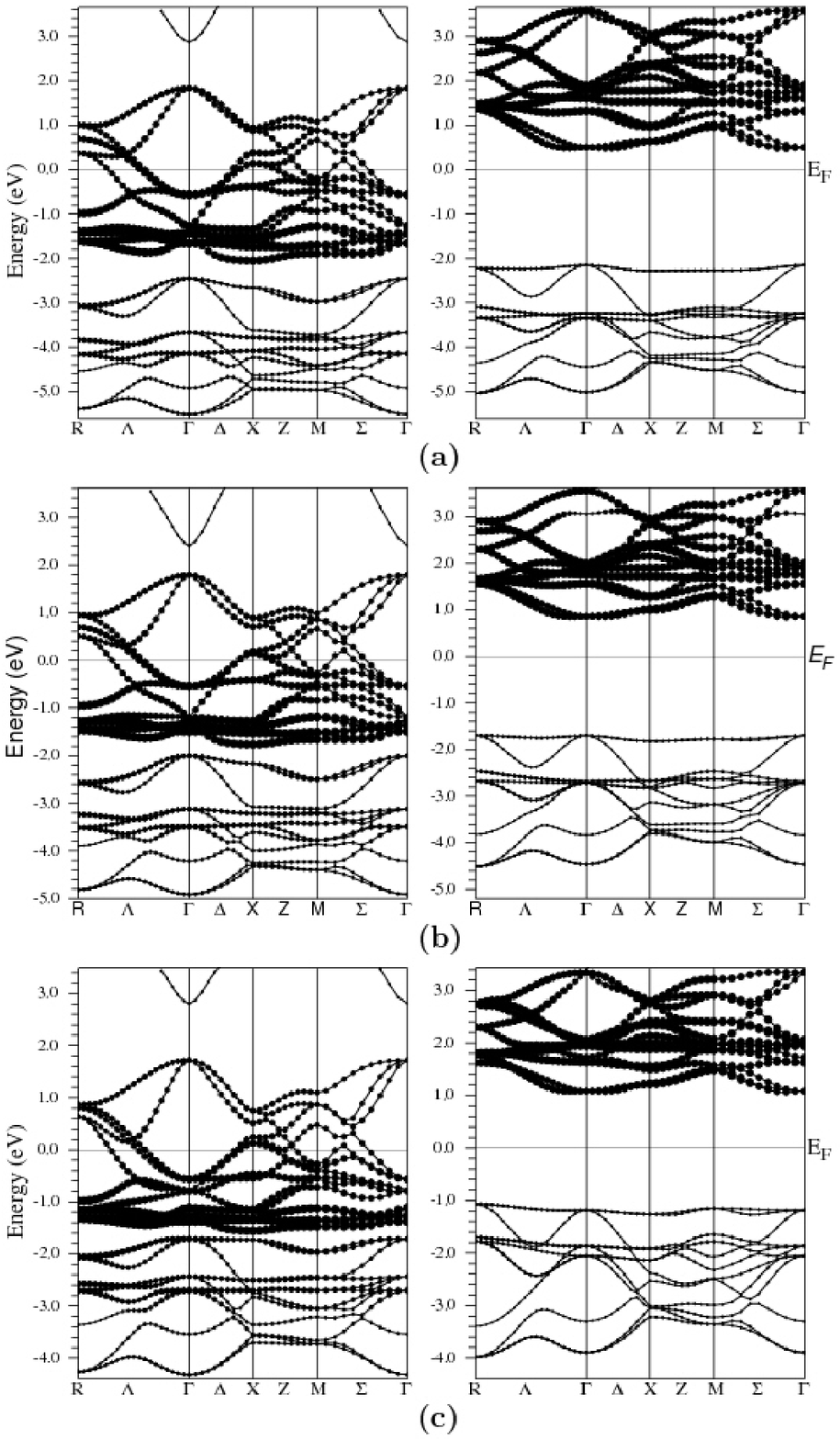}
\vspace{0mm} \caption{Spin-dependent energy band structures (EBs)
of the three sulvanite Cr$_3$$X$$_4$ phases for $X$ = S (a), Se
(b), and Te (c). The left part in each of the three cases is the
majority-spin bands and the right part the minority spin. The
solid line with dots describes the EB structure along the
high-symmetry points and the dot diameter is proportional to the
Cr $d$ weight at that point.} \label{fig3}
\end{center}
\end{figure}

The spin-dependent DOSs of the three sulvanite Cr$_3$$X_4$ ($X$=S,
Se, and Te) phases are presented in Fig. 2. For the convenience of
description, we label the four different sets of the energy bands
as A, B, C, and D, respectively. For each of the three $X$ cases,
there is a narrow gap between A and B in the majority-spin (MAS)
channel and a wide gap between C and D in the minority-spin (MIS).
It is clear that there is a gap across the Fermi level in the MIS
channel for each of all the three cases and therefore all the
three are half-metallic ferromagnets. In addition to the total
DOS, partial DOSs projected in the muffin-tin spheres of Cr and
$X$ atoms and in the interstitial region are presented too. It is
clear that the B and D bands are originated from Cr-d states and
the A and C bands mainly have the $X$-p (S, Se, or Te) character.

Presented in Fig. 3 are the spin-dependent energy bands, according
to the corresponding DOS plots in Fig. 2, for $X$=S, Se, and Te.
There are twelve bands in both A and C. They result from the fact
that we have four $X$ (S, Se, or Te) atoms in the sulvanite
Cr$_3$$X$$_4$ unit cell and each of the four has three p orbitals.
There are fifteen bands in both B and D because we have three Cr
atoms here and each of the three has five d orbitals.  The $e_g$
bands are lower than the $t_{2g}$ ones. The Cr-s state is put to
higher energy than the B and D bands in both of the MAS and MIS
channels. This is in contrast with zb-Cr$X$ phases whose Cr-s
states are merged with the Cr-d $t_{2g}$ states. We have also
calculated charge and spin densities in three typical planes. Our
results show that the charge density from the A and C bands is
distributed mainly in the neighborhoods of the $X$ and Cr atoms,
but the spin density is almost limited to Cr and $X$ atoms only.
The filled part of the B bands are substantially localized near Cr
and $X$ atoms. When $X$ changes from S to Te, the $X$-p character
becomes less and less in the electron density from the filled B
bands because the ionic radius becomes larger and larger.

\begin{table}[!htbp]\begin{center}
\caption{The magnetic moments ($M$), half-metallic gaps ($G_{\rm
HM}$), magnetic energy differences ($\Delta_E$), formation heats
($H_{\rm Form}$), and the elastic moduli ($B$, $C^\prime$, and
$C_{44}$) of the three Cr$_3$$X_4$ phases, compared with the
zb-Cr$X$ phases (in parentheses).}\label{table2}
\begin{tabular*}{0.49\textwidth}{@{\extracolsep{\fill}}cccc}
 \hline\hline
Name & Cr$_3$S$_4$(CrS) & Cr$_3$Se$_4$(CrSe) & Cr$_3$Te$_4$(CrTe)\\
\hline
$M$ ($\mu_B$)  & 10 (4) & 10 (4)  & 10 (4)  \\
$G_{\rm HM}$ (eV) & 0.50 (0.07) & 0.83 (0.61)  & 1.05 (1.00)\\
$\Delta_E$ (eV) & 0.57   & 0.56  & 0.11 \\
$H_{\rm Form}$ (eV) & -0.148 &  -0.114 & -0.302  \\
$B$ (GPa)       & 72.9 (63.2) & 57.7 (59.5) & 44.1 (45.9)\\
$C^\prime$ (GPa) & 8.8 & 5.9 (5.6) & 4.8 (5.5)\\
$C_{44}$ (GPa) & 34.2 & 35.9 (50.7) & 32.9 (36.4)\\
 \hline\hline
\end{tabular*}\end{center}
\end{table}

We summarize in Table \ref{table2} the total moments per formula
unit ($M$), half-metallic gaps ($G_{\rm HM}$)\cite{lbghm}, and the
energy differences per formula unit ($\Delta_E=E_{\rm AF}-E_{\rm
FM}$) for all the three sulvanite Cr$_3$$X_4$ phases ($X$=S, Se,
and Te), where $E_{\rm AF}$ and $E_{\rm FM}$ are the total
energies for the AF and FM spin configurations. Here we define
$G_{\rm HM}$ as the smaller of $E_{cb}$ and $E_{vt}$, where
$E_{cb}$ is the bottom energy of the minority-spin conduction
bands with respect to the Fermi level and $E_{vt}$ the absolute
value of the top energy of the minority-spin valence bands. Our
GGA $G_{\rm HM}$ values should be rough estimates for the minimal
energies for spin flip excitations. We obtain 10$\mu_B$ for the
magnetic moment per formula unit in all the three cases. This is
because we have ten Cr d electrons to contribute to the moment
after eight of the eighteen Cr electrons of the three Cr atoms are
bonding with the sixteen $X$ p electrons of the four $X$ atoms.
The half-metallic gaps are larger than the corresponding zb-Cr$X$
phases because the Fermi levels move toward lower energy level due
to the fact that there are one less Cr atom for Cr$_3$$X_4$,
compared with the cubic zb unit cell including four Cr and $X$
atoms. The $\Delta_E$ results, from 0.11 to 0.57 eV, show that our
FM solutions are very robust against possible AF fluctuations.

In addition, we present the formation heats per formula unit
($H_{\rm Form}$), the elastic moduli ($B$, $C^\prime$, and
$C_{44}$) of the the three Cr$_3$$X_4$ phases in Table
\ref{table2}. We calculate the formation heats to investigate the
stability of the Cr$_3$$X_4$ with respect to the zb-Cr$X$ phases.
For this reason, $H_{\rm Form}$ is defined as $E({\rm
Cr}_3X_4)-4E({\rm zb\!-\! Cr}X)+E({\rm Cr})$, where $E(f)$ is the
total energy of the formula $f$. The negative values of $H_{\rm
Form}$ for all the three cases mean that the sulvanite Cr$_3$$X_4$
phases can be synthesized more easily than the corresponding zb
phases, since zincblende CrTe has been synthesized in the form of
epitaxial thin films\cite{crte}. The $B$ and $C_{44}$ values of
the sulvanite phases are similar to those of the corresponding
zincblende phases. The key tetragonal moduli, $C^\prime$, are 8.8,
5.9, and 4.8 GPa for the three cases. The latter two $C^\prime$
values are comparable with those of the corresponding zincblende
phases. Therefore, these calculated results show that the three
sulvanite phases are mechanically stable.

The nonstoichiometry plays key roles in the sulvanite Cr$_3$$X_4$.
A Cr atom in the zb-Cr$X$ contributes a moment of 4$\mu_B$, but
three Cr atoms yield 10$\mu_B$, not 12$\mu_B$, in the sulvanite
Cr$_3$$X_4$ because two more Cr electrons are needed to fill all
the p orbitals of four $X$ atoms. Effectively, the average valence
of a Cr atom is equivalent to $\frac 83$$+$ (higher than $2+$) and
the enhanced Cr valence makes the formation heat negative and
hence enhances the stability of the Cr$_3$$X_4$ with respect to
the zb-Cr$X$. In zb-Cr$X$, Cr-d$_{t2g}$ states are hybridized
substantially with $X$ p ones, and Cr-d$_{eg}$ are quite isolated.
For the Cr$_3$$X_4$, Cr-d$_{x^2-y^2}$ states remain isolated, and
the other Cr d states are hybridized more strongly with the $X$ p
ones, which makes Cr s state higher in energy.

In summary, we show that the three nonstoichiometric cubic binary
chromium chalcogenides are stable half-metallic ferromagnets with
wide half-metallic gaps on the basis of systematic
state-of-the-arts DFT calculations. Our calculated results
indicate that the sulvanite Cr$_3$S$_4$, Cr$_3$Se$_4$, and
Cr$_3$Te$_4$ are better than or approximately equal in both
structural and ferromagnetic stability than zincblende CrS, CrSe,
and CrTe, respectively. The structural stability against crystal
deformations and the negative formation heats (-0.114$\sim$-0.302
eV per formula unit) mean that they can be synthesized, at least
as epitaxial thin films, more easily than the corresponding
zincblende phases. We attribute the structural and ferromagnetic
stability and the better half-metallicity to the higher Cr
chemical valence, $2.667+$, in the sulvanite structures than $2+$
in the corresponding zincblende ones. These findings will open
doors for much more high-performance spintronic materials
compatible with current semiconductor technology.

\acknowledgments This work is supported  by Nature Science
Foundation of China (Grant Nos. 10874232 and 10774180), by the
Chinese Academy of Sciences (Grant No. KJCX2.YW.W09-5), and by
Chinese Department of Science and Technology (Grant No.
2005CB623602).

\end{document}